\documentclass[twocolumn,pre,aps,showpacs,preprintnumbers,amsmath,amssymb,superscriptaddress]{revtex4}

\usepackage{epsfig}

\usepackage{graphics}
\usepackage{dcolumn}
\usepackage{bm}

\newcommand{\avg}[1]{\langle #1 \rangle}

\begin{document}


\title{Hybrid recommendation methods in complex networks}

\author{A. Fiasconaro}
 \email{A.Fiasconaro@qmul.ac.uk}
 \affiliation{School of Mathematical Sciences, Queen Mary
University of London. Mile End Road, London E1 4NS, UK}

\author{M. Tumminello}
\affiliation{Dipartimento di Scienze Statistiche e Matematiche,
Universit\`a di Palermo, Viale delle Scienze, 90128 Palermo,
Italy}

\author{V. Nicosia}
 \affiliation{School of Mathematical Sciences, Queen Mary
University of London. Mile End Road, London E1 4NS, UK}

\author{V. Latora}
 \affiliation{School of Mathematical Sciences, Queen Mary
University of London. Mile End Road, London E1 4NS, UK}

 \affiliation{Dipartimento di Fisica ed Astronomia,
Universit\`a di Catania and INFN, I-95123 Catania, Italy}

 \author{R.N. Mantegna}
 \affiliation{Center for Network Science, Central European University, Nador 9 ut., H-1051, Budapest, Hungary}
\affiliation{Department of Economics, Central European University, Nador 9 ut., H-1051, Budapest, Hungary}
\affiliation{Dipartimento di Fisica e Chimica, Universit\`a di
Palermo, Viale delle Scienze, Edif. 18, I-90128, Palermo, Italy}

\date{\today}

\begin{abstract}
We propose here two new recommendation methods, based on the
appropriate normalization of already existing similarity measures, and
on the convex combination of the recommendation scores derived from
similarity between users and between objects. We validate the proposed
measures on three relevant data sets, and we compare their performance
with several recommendation systems recently proposed in the
literature. We show that the proposed similarity measures allow to
attain an improvement of performances of up to 20\% with respect to
existing non-parametric methods, and that the accuracy of a
recommendation can vary widely from one specific bipartite
network to another, which suggests that a careful choice of the most
suitable method is highly relevant for an effective recommendation on
a given system. Finally, we studied how an increasing presence of random links in the network affects the recommendation scores, and we found that one of the two recommendation algorithms introduced here can systematically outperform the others in noisy data sets.
\end{abstract}


\maketitle

\section{Introduction}

The increasingly important role played by information technology and
by the ubiquity and success of web-based retail shops is rapidly
transforming our lives and buying patterns, and is producing a huge
quantity of detailed data sets about customers' preferences and
habits. The availability of such data sets has made possible to study
in a quantitative way how people select items in several different
scenarions such as, for instance, how they choose movies to watch,
books to read, or food to eat. In most of the cases, the number of
different items available on an online retail shop is so large that it
is extremely difficult to have a clear idea of the specific products
that would better fit the taste of each customer. Hence the necessity
to devise intelligent automatic systems to provide useful
recommendations, based for instance on the knowlegde about previous
purchases made by users. Given its practical importance, the study of
recommendation systems is nowadays a very active research topic,
with relevant contributions from different fields including computer science, economics,
sociology, complex networks, and
engineering~\cite{2005IEEEAdomavicius,2012PRLu}.

The natural framework to represent selection or purchasing patterns is
by means of a bipartite graph, namely a graph consisting of two
distinct classes of nodes (respectively associated to users and
objects) in which two nodes belonging to different classes are
connected by an edge if the corresponding user has chosen or purchased
that particular object. Within this framework, a recommendation is no
more than the suggestion of (a relatively small) set of objects to
which a specific user might be interested, and corresponds to a set of
new potential edges in the bipartite graph. In many cases, an object
is recommended to a user based on her similarity with other users, so
that the definition of appropriate similarity measures is crucial for
the development of efficient {\em personalized recommendation
  systems}. Various recommendation systems and algorithms have been
proposed over the years, such as the collaborative filtering
(CF)~\cite{1992ACMGoldberg,2007Schafer,2001Sarwar}, methods based on
diffusion across the user-object
network~\cite{2007PRLZhang,2007PREZhou,2009PNASZhou,2009NJPZhou,2011PRELu},
and hybrid (parametric) combinations of different
algorithms~\cite{2011PRELu,2009PNASZhou,2013PONEQiu,2014arXivZhu}.  In
most of the cases, the quantification of similarity between two users
is based on the number of objects which have been chosen by both users
in the past. However, it is also possible to define a similarity
between two objects based on the number of users who have chosen them.
Not always in the literature recommendations based on user similarity
have been properly distinguished from those based on object
similarity, and the predictions provided by these two types of
recommendation systems have been usually compared while discarding the
different nature of the similarity measures involved.

The question of which similarity measure is the most reliable in
providing tailored and accurate recommendations is still a matter of
open debate, and it is not clear yet how to choose one similarity
definition or another for a specific recommendation task. This paper
provides a contribution in this direction.
In particular, we focus here on the duality user-similarity versus
object-similarity, showing that it is possible to improve the quality
of recommendation by making a combined use of the two classes of
similarity.  We start by proposing two new definitions of similarity
based on heuristic arguments, and then we compare the accuracy of
recommendations based on these definitions with the accuracy of other
methods proposed in the literature. The first measure we propose takes
into account the popularity of objects and the heterogeneity of user
selection patterns, while the second one is based on the concept of
Pearson correlation~\cite{1895PRSLpearson} generalized to the case of
binary vectors. We then show that any definition of similarity between
users induces a definition of similarity between objects, and vice
versa. This fact actually increases the number of different possible
similarity definitions, and allows to consider recommendation methods
which combine together user-similarities with object-similarities. We
also test the robustness of different similarity measures against the
presence of noise, by adding an increasing percentage of random edges
to the actual bipartite graph, and we show that the measure based on
generalised Pearson correlation proposed in the paper is able to
filter out noise more effectively than most of the other existing
similarity measures.

The paper is organized as follows. In Section~\ref{Measures} we review
different similarity measures proposed in the literature, we introduce
two new definitions of similarity, and we show how to associate a
recommendation score to an object starting from a given similarity
definition. In Section~\ref{Validation} we provide a brief description
of three data sets corresponding to to user/object associations in
different contexts, and we validate the performance of different
recommendation strategies on the corresponding bipartite networks. In
Section~\ref{Sec:Specular} we show how a measure of similarity between
users can be transformed into an analogue measure of similarity
measure between objects. We then investigate recommendation methods
which combine the recommendation scores obtained from similarities
between users and similarity between objects. Finally, in Section~\ref{Randomization} we study how the presence of spurious
information in the data sets can affect the performance of
recommendation, and we show that some recommendation strategies
actually perform better than others in noisy data sets.

\section{Measures of similarity}
\label{Measures}

Let us consider a set of $N$ users and $M$ objects, where each user
$i$ is associated to a subset of the $M$ objects she has expressed a
preference for. This is for instance the case of users buying objects
from an online retail shop, where each user is associated to each of
the items she has bought from that website. Such systems can be
naturally represented by means of bipartite graphs, where users and
objects are considered as two distinct classes of nodes. A bipartite
graph can be described by a $N \times M$ adjacency matrix $A$ whose
entry $a_{i,\alpha}$ is equal to 1 if and only if user $u_i$ is
associated to object $o_\alpha$ (for instance, because $u_i$ has
bought that object), and $a_{i,\alpha} = 0$ otherwise. Notice that in
a bipartite graph each edge always connects one user with one
object. In the following, Latin subscripts are associated to users,
whereas Greek ones are associated to objects. The total number of
objects collected by a user $u_i$ is equal to the number of edges
incident on the corresponding node $i$ of the graph, i.e. to the
degree $k_i=\sum_{\alpha}a_{i,\alpha}$. Similarly, the degree of
object $o_{\alpha}$, defined as $k_{\alpha}=\sum_{i} a_{i,\alpha}$, is
equal to the total number of users that have collected that object.

Within this framework, making a \textit{recommendation} for user $u_i$
corresponds to compiling a list of objects which have not already been
chosen (or bought) by user $u_i$ but to which $u_i$ might be
interested. In other words, a recommendation is just a proposal of new
potential edges of the bipartite graph whose one endpoint is node $i$.
The main hypothesis on which almost all recommendation systems rely is
that the set of objects actually collected (or bought) by user $u_i$
represents a sample of her tastes and preferences, and can therefore
be used to compile a profile of user $u_i$ and to predict which kind of
objects $u_i$ might be interested. Consequently, each recommendation
systems relies on some measure of similarity. In general, it is
possible to define a similarity $s^{\rm u}_{i,j}$ for the (ordered)
pair of users $u_i$ and $u_j$ and also a similarity $s^{\rm
  o}_{\alpha,\beta}$ between the pair of objects $o_{\alpha}$ and
$o_{\beta}$, and various different definitions have been proposed in
the literature~\cite{2011PALu}.

\textit{Similarity measures. ---} A very simple way of quantifying the
similarity between user $u_i$ and user $u_j$ is by counting the number
of objects $n_{i,j}$ that they have in common:
\begin{equation*}
  n_{i,j}=\sum^{M}_{\alpha=1}{a_{i,\alpha}
  a_{j,\alpha}}
\end{equation*}
One of the limitations of $n_{i,j}$ is that it doesn't take into
account the differences in the total number of objects collected by
each user.  This problem can be somehow alleviated by using the
so-called Jaccard similarity~\cite{1912NPjaccard}, defined as the
ratio between the number of items collected by both users $u_i$ and
$u_j$, and the sum of the degrees of the two users:
 \begin{equation} s^{\rm u, J}_{i,j} =
\frac{\sum^{M}_{\alpha=1}{a_{i,\alpha} a_{j,\alpha}}}
{(k_i+k_j)}=\frac{n_{i,j}} {(k_i+k_j)}.
 \label{s-J}
 \end{equation}

Another widely used similarity measure is the one of the
\emph{collaborative filtering} (CF) approach, which is defined as
 \begin{equation} s^{\rm u, CF}_{i,j} =
\frac{\sum^{M}_{\alpha=1}{a_{i,\alpha} a_{j,\alpha}}}
{\min\{k_i,k_j\}}.
 \label{s-Min}
 \end{equation}
In Eq.~(\ref{s-Min}), the similarity measure is proportional to the
number of objects $n_{i,j}$ users $u_i$ and $u_j$ have in common, and
inversely proportional to the smallest of the two degrees, i.e. to
$\min \{k_i,k_j\}$. In this way, if user $u_i$ has collected exactly
one object, which has also been collected by $u_j$ who instead has
degree $k_j\gg 1$, then $s^{\rm u, CF}_{i,j}=1$, i.e. the similarity
between two users is effectively determined by the user with the
smallest degree.

The Jaccard and the collaborative filtering similarity do not take into
account another type of heterogeneity, namely the fact that not all
objects have the same popularity. Intuitively, objects that have been
collected by a relative large number of users (in a limiting case, by
all users), do not provide useful information for a personalised
recommendation, for the simple reason that they are common to too many
users, and therefore the fact that one user has collected them does
not tell much about her tastes. Hence, it might be a good idea to
discount the contribution of an object to the similarity between two
users by a function of the degree of the object.

The so called Network-Based Inference (NBI) recommendation
method~\cite{2007PREZhou} is based on a measure of similarity which
takes into account the heterogeneity of users and objects (this
recommendation strategy is also called \textit{probabilistic
  spreading} in a subsequent paper~\cite{2009PNASZhou}). In this case,
the similarity measure is defined as
\begin{equation}
s^{\rm o, NBI}_{\alpha,\beta} = \frac{1}{k_{\beta}} \sum^{N}_{l=1}
\frac{a_{l,\alpha} a_{l,\beta}} {k_l}
 \label{s-NBI}
\end{equation}
This is a similarity between objects, where the contribution of the
user $u_l$ which collects the two objects $\alpha$ and $\beta$ is
discounted by the degree of that user $k_l$, and the whole sum is
divided by the degree of one of the two objects, according to the
\emph{resource-allocation} procedure defined by the authors in
Ref.~\cite{2007PREZhou}.

It is worth noting that this definition of similarity, like the
analogous one $s^{\rm o, HeatS}_{\alpha,\beta} = \frac{1}{k_{\alpha}}
\sum^{N}_{l=1} \frac{a_{l,\alpha} a_{l,\beta}} {k_l}$ investigated in
\cite{2007PRLZhang,2009PNASZhou}, is asymmetric, meaning that $s^{\rm
  o, NBI}_{\alpha,\beta} \neq s^{\rm o, NBI}_{\beta,\alpha}$, and
$s^{\rm o, HeatS}_{\alpha,\beta} \neq s^{\rm o,
  HeatS}_{\beta,\alpha}$. Though asymmetry is not in general an issue
for the recommendation task, it has been shown that better performance
can be achieved by using symmetrized versions of these
measures~\cite{2009PNASZhou,2014arXivZhu}. Nevertheless, NBI has
proved to be a quite reliable recommendation method, and in the
following we will consider it as a reference to quantify the
effectiveness of the recommendation strategies we propose.

The two new recommendation methods we propose in this paper are based
on symmetric similarity measures. Specifically, the first measure we propose is
 \begin{equation}
s^{\rm u,MDW}_{i,j} = \frac{1}{\max\{k_i,k_j\}}
  \sum^{M}_{\alpha=1, k_{\alpha}>1}\frac{{a_{i,\alpha} a_{j,\alpha}}} {k_{\alpha}-1}.
 \label{s-MDW}
 \end{equation}
We call it Maximum Degree Weighted (MDW) similarity because the
total number of objects collected by both $u_i$ and $u_j$ is weighted
by the maximum of the degrees of the two users. Moreover, the
contribution of object $o_{\alpha}$ is weighted by its degree
$k_{\alpha}$. We note here that some recent studies have investigated
the effect of a tunable power-law function of the degree, i.e. of
similarity measures in which the contribution due to object
$o_{\alpha}$ is divided by
$\left(k_{\alpha}\right)^a$~\cite{2009PALiu,2012EPJBGuo,2007PRLZhang}.

In the following, we will briefly explain the rationale behind Eq.~(\ref{s-MDW}). First
of all, the contribution to the similarity measure of each object
collected by both users is weighted with the degree of the object. In
this way, popular objects will provide smaller contributions to the
similarity between users. In particular, the value $k_{\alpha}-1$ in
the denominator allows to obtain a maximum contribution to similarity
(exactly equal to 1) if and only if $a_{i,\alpha} a_{j,\alpha} =1$ and
$k_{\alpha}=2$. This takes into account the very special case in which
$u_i$ and $u_j$ are the only two users who have collected a certain
object.  Secondly, the similarity measure is divided by the maximum of
the degrees of the two users. As we explained above, this choice
allows to properly take into account the existing heterogeneity in the number
of selection made by each user. For instance, if we consider the
similarity defined in Eq.~(\ref{s-Min}) and we assume that users $u_i$
and $u_j$ have degree $k_i=1$ and $k_j=100$ and have exactly one
object in common, then the contribution of that object to the
similarity between the two users would be equal to $1$.  However, the
contribution of the only object in common between $u_i$ and $u_j$ is
equal to $1$ also when $k_i=1$, and $k_j=2$, despite one would argue
that in the latter case the two users are more similar than in the
former.

By dividing for the maximum of the degrees of the two users, the
similarity measure given in Eq.~(\ref{s-MDW}) assigns a higher value
of similarity to the two users in the latter case (when $k_i=1$ and
$k_j=2$ we get $s^{\rm u}_{i,j}=1/2$) than in the former case (i.e.,
when $k_i=1$ and $k_j=100$, for which we obtain $s^{\rm
  u}_{i,j}=1/100$).

A second similarity measure we propose here is based on the Pearson
correlation coefficient between binary vectors, and is defined as
follows~\cite{2011PoneAGing}:
 \begin{equation}
 s^{\rm u,BP}_{i,j} =  \frac{n_{i,j}-k_ik_j/M}
{\sqrt{k_i(1-k_i/M)k_j(1-k_j/M)}}.
 \label{s-P}
 \end{equation}
This measure, which is denoted in the following as Binary Pearson
(BP) similarity, is based on the fact that the $i^{\rm th}$ row $\mathbf{a}_{i}$ of the 
adjacency matrix $A$ represents the profile vector of user $u_i$, i.e.
the set of objects selected by the user. If we have two users, $u_i$ and $u_j$, 
who have collected $k_{\alpha}$ and $k_{\beta}$
objects in total, respectively, and we consider the corresponding profile
vectors $\mathbf{a}_i$ and $\mathbf{a}_j$, then we have:
\begin{eqnarray}
\avg{{a_i}}& =& \frac{1}{M}\sum_{\alpha=1}^{M} a_{i,\alpha} = \frac{k_i}{M}\\
\avg{{a_i}^2}& =& \frac{1}{M}\sum_{\alpha=1}^{M} a_{i,\alpha}^2 = \frac{1}{M}\sum_{\alpha=1}^{M} a_{i,\alpha} =\frac{k_i}{M},\\
\avg{{a_i}{a_j}}& =& \frac{1}{M}\sum_{\alpha=1}^{M} a_{i,\alpha}\,a_{j,\alpha} =\frac{n_{i,j}}{M},
\end{eqnarray}
where $n_{i,j}$ is the number of objects collected by both $u_i$ and
$u_j$. Therefore the Pearson's correlation coefficient between the two
vectors is:
\begin{eqnarray}
s^{\rm u,BP}_{i,j} &=& \frac{\avg{{a_i} {a_j}} -
  \avg{{a_i}}\,\avg{{a_j}}}{\sqrt{(\avg{{a_i}^2}
    - \avg{{a_i}}^2)\,(\avg{{a_j}^2} -
    \avg{{a_j}}^2)}} \nonumber\\  & = &
\frac{\displaystyle\frac{n_{i,j}}{M}-\frac{k_i\,k_j}{M^2}}{\sqrt{\left(\frac{k_i}{M}-\frac{k_i^2}{M^2}\right)\,\left(\frac{k_j}{M}-\frac{k_j^2}{M^2}\right)}}=\nonumber\\ &
= & \frac{n_{i,j}-k_ik_j/M} {\sqrt{k_i(1-k_i/M)k_j(1-k_j/M)}}.
 \end{eqnarray}
which is identical to Eq.~(\ref{s-P}).  This similarity measure has
some remarkable properties. First of all, it is invariant with respect
to a rescaling of the system, i.e. with respect to a
transformation $k_i \rightarrow q\,k_i$, $k_j \rightarrow q\,k_j$,
$n_{i,j} \rightarrow q\,n_{i,j}$, and $M \rightarrow q \, M$, where
$q$ is a positive integer. Furthermore, $s^{\rm u,P}_{i,j}$ can be
interpreted in terms of the hypergeometric distribution
$H(X,k_i,k_j,M)$. Indeed, the mean value of $H(X,k_i,k_j,M)$ is
$m_{i,j}=k_i\,k_j/M$ and the variance is
$\sigma_{i,j}^2=\frac{1}{M-1}k_i(1-k_i/M)k_j(1-k_j/M)$. Therefore
$s^{\rm u,P}_{i,j}$ is proportional to the standard score $z_{i,j}^H$
associated with observation $n_{i,j}$ according to the hypergeometric
distribution \cite{2014Hatzopoulos}:

\begin{equation*}
  s^{\rm u,P}_{i,j}=\frac{1}{\sqrt{M-1}}\,\frac{n_{i,j}-m_{i,j}}{\sigma_{i,j}}=\frac{1}{\sqrt{M-1}}\,z_{i,j}^H.
\end{equation*}

This equation reveals that $s^{\rm u,P}_{i,j}$ is conceptually
different from all the other similarity measures introduced above.
Indeed, according to $s^{\rm u,P}_{i,j}$, the similarity between two
users does not depend only on the number of objects selected by both
users, $n_{i,j}$, but it depends on the difference between $n_{i,j}$
and the number of shared objects that is expected under the hypothesis
that the two users have picked the objects at random. Therefore
$s^{\rm u,P}_{i,j}$ can also take negative values, and this fact
influences the way in which a personalized recommendation value is
obtained, as discussed in the reminder of this section.

\textit{Constructing recommendation lists. ---} Once we have assigned
a similarity value to each possible pair of users in the system, using
a certain similarity measure, we need an algorithm to construct a
recommendation list, i.e. a list of suggested objects which have not
been yet collected by a certain user $u_i$. The simplest way of
constructing a recommendation list is the \emph{Global Ranking Method}
(GRM). It consists in creating a user recommendation list by
considering all the objects not collected by user $u_i$ in decreasing
order of their degree.  This method is not personalized, except for
the fact that objects already collected by that user are excluded from
the corresponding list.

A more effective and widely used basic procedure is the
\textit{Collaborative Filtering} (CF), which is based on the
similarity measure given in Eq.~(\ref{s-Min}). The similarity scores
between user $u_i$ and all the other users in the system are used to
construct a personalized recommendation value $v^{\rm u}_{i,\alpha}$,
that is an estimation of how much user $u_i$ might be interested on
object $o_{\alpha}$:
 \begin{equation}
  v^{\rm u}_{i,\alpha} = \frac{\sum^{N}_{l=1, l\neq i} {s^{\rm u}_{i,l} \cdot a_{l,\alpha}}}
{\sum^{N}_{l=1, l\neq i} {\left | s^{\rm u}_{i,l} \right |} }.
 \label{eq:vu}
 \end{equation}
 The presence of the absolute value at the denominator is not necessary for many of the similarities presented above, being their values always positive. The only exception is the BP similarity, which may take both positive and negative values, thus requiring a proper normalization to avoid possible divergences.
 In the NBI recommendation method, the recommendation value $f^{\rm
   o}_{i,\alpha}$ is defined in a quite different way
 \begin{equation}
  f^{\rm o}_{i,\alpha} = \sum^{M}_{\beta=1} {s^{\rm o, NBI}_{\alpha,\beta} \cdot a_{i,\beta}}.
 \label{eq:fo}
 \end{equation}
In fact the computation of the recommendation value includes self
similar terms ($s^{\rm o}_{\alpha,\alpha}$) which are not taken into account in Eq.~(\ref{eq:vu}),
and does not include any additional normalization factor.

\section {Data sets and validation}
\label{Validation}

We considered three classical data sets of user-object
associations, namely the MovieLens
(http://www.grouplens.org/node/73) database, where $N$ users have
rated the $M$ movies, the Jester Jokes database
(http://eigenstate.berkeley.edu/dataset/) where we find records of
users who have rated jokes, and the Fine Foods database
(https://snap.stanford.edu/data/web-FineFoods.html), containing
Amazon reviews of Fine foods. In all these databases users have
rated items with an ordinal attribute. In our study we will
perform recommendation procedures on the adjacency matrix of the
corresponding bipartite network. For this reason, according to a
similar choice done in other studies (see
Ref.~\cite{2007PREZhou}), we assume that a user has collected an
object if and only if he has rated the object with a score higher
or equal to a certain preselected threshold. In
Table~\ref{table:DB} we report some information about the size of
each data set and the values of the thresholds considered for the
definition of the corresponding bipartite network.

\begin{figure}
 \centering
\includegraphics[angle=-90,width=8.5cm]{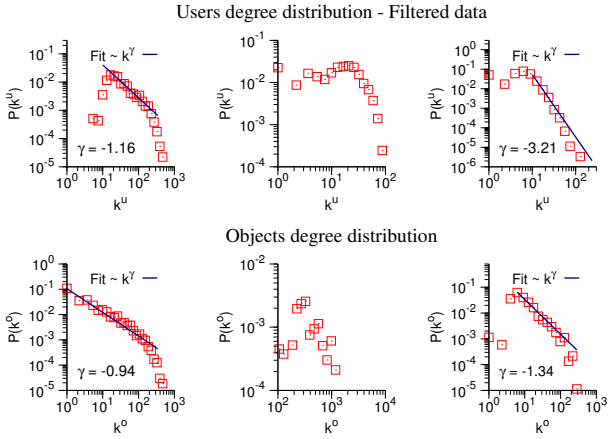}
\caption{(Color online) Degree distributions of users (top panels) and
  objects (bottom panels) of the three data sets, respectively
  MovieLens (left), Jester Jokes (middle), and Fine Food (right).}
 \label{DB_f}
 \end{figure}

\textit{Distributions of similarity values. ---} As a preliminary
investigation, we evaluated the heterogeneity of the degree of users
and objects in the three databases. In Fig.~\ref{DB_f} we show the
degree distributions for the three databases used. With the only
exception of the degree distribution of jokes (objects) in the Jester
Jokes database, all the distributions exhibit relatively broad
tails. This suggests that similarity measures which properly take into
account degree heterogeneities should indeed provide better
recommendations.
\begin{table}[ht]
\caption{Summary statistics of the three databases used in the paper}
\centering
\begin{tabular}{l c c c c c} 
\hline\hline 
           & $N_{links}$ &  N x M   &  Rating  &  Thre-  & $N_{links}$\\
           &             &  (used)  &  range   &  shold    &  Filtered  \\ [0.2ex] 
\hline 
 Movielens  &   100K  & 943 x 1,681 & [1, 5]  &  3   &   90K \\
 Jester Jokes &   141K   & 2,000 x 100  & [-10, 10]  & 0 &   57K  \\
 Fine Foods &   95K & 2,000 x 3,317 &  [1, 5] &  3  &   83K  \\
\hline
\end{tabular}
\label{table:DB}
\end{table}

\begin{figure}
 \centering
\includegraphics[angle=-90,width=8.5cm]{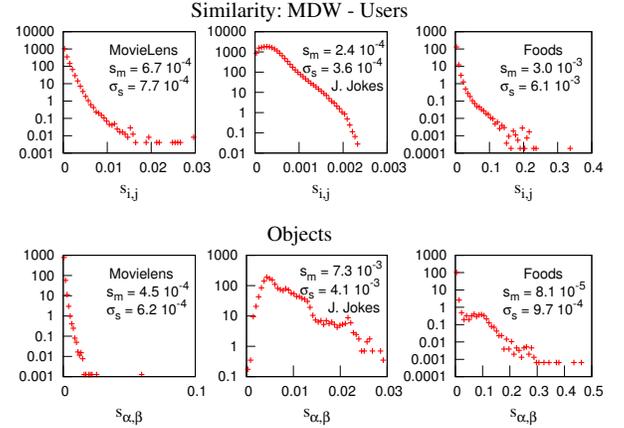}
\caption{Distribution of the MDW similarity values for the three
  databases. The top panels show the probability density functions of
  users' similarity whereas the bottom panels show the probability
  density functions of object similarity.  Left, middle, and right
  columns refer to MovieLens, Jester Jokes, and Fine Food databases
  respectively.}
 \label{f:Sim_IMUW}
 \end{figure}

\begin{figure}
 \centering
\includegraphics[angle=-90,width=8.5cm]{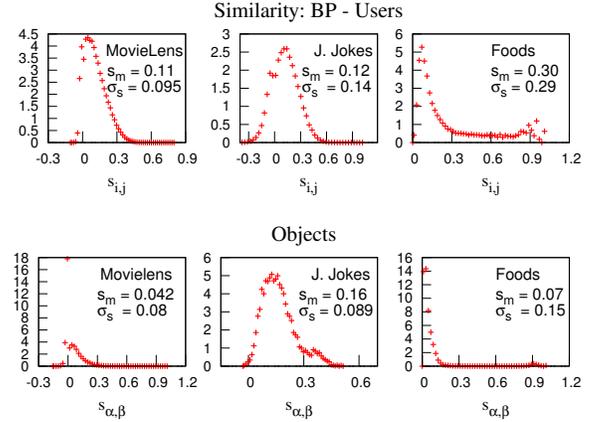}
\caption{Distribution of the binary Pearson similarity values for the
  three databases. The top panels show the probability density
  functions of users' similarity whereas the bottom panels show the
  probability density functions of object similarity.  Left, middle,
  and right columns refer to MovieLens, Jester Jokes, and Fine Food
  databases respectively.}
 \label{f:Sim_P}
 \end{figure}

As a matter of fact, different similarity measures produce different
distribution of similarity scores.  In Fig.~\ref{f:Sim_IMUW} and
Fig.~\ref{f:Sim_P} we show the probability density functions of the
MDW and BP similarities measures on the three investigated
databases. The top panels of each figure report the distribution of
user similarities, whereas the bottom panels correspond to object
similarity. It is worth noting that the profile of the probability
density functions is strongly dependent on the similarity measure
adopted and is qualitatively different in the three data sets.

\textit{Validation. ---} In order to compare the performance of the
proposed similarity measures with those of other existing similarity
definitions, we split each data set into two sections. Starting from
the adjacency matrix $A=\{a_{i,\alpha}\}$ representing all the
user-object associations in a data set, we considered a subgraph
$T=\{t_{i,\alpha}\}$ to be used to compute the similarity scores and
recommendation lists for all the users (the so-called \textit{training
  set}), while the remaining subgraph $W = \{a_{i,\alpha}\} \setminus
\{t_{i,\alpha}\} = \{w_{i,\alpha}\}$ was used for
\textit{validation}. The recommendation lists obtained from the
training sets are compared with the object selections included in the
validation set, in order to check whether users have actually
collected objects which are ranked high in their recommendation
lists. In the following we report the results corresponding to
training sets containing $90\%$ of all the edges of each data set,
chosen at random, while the validation sets consist of the remaining
$10\%$ of edges. Qualitatively similar results were obtained for
different compositions of the training and validation sets.

A basic measure to quantify the performance of a recommendation method
is the \textit{rank quality index} $r$, which is computed as the
average quality of recommendation over all the users of the data
set. For each user $u_i$ we define the quality of the recommendation
$r_i$ provided to $u_i$ as the average of the ratio
\begin{equation}
  r_{i,\alpha} = w_{i,\alpha}\frac{L_{i,\alpha}}{M-k_i}.
  \label{eq:rialpha}
\end{equation}
computed over all the objects in the recommendation list of user $u_i$
which have actually been selected by $u_i$ in the validation set. Here
$L_{i,\alpha}$ is the position of $o_{\alpha}$ in the recommendation
list of $u_i$, where $L_{i,\alpha}=1, 2, \ldots, $ if the object
$o_{\alpha}$ is ranked first, second,$\dots$ etc. in the
recommendation list of $u_i$. Consequently, better recommendations are
associated to smaller values of $r_{i} = \sum_{\alpha} r_{i,\alpha}$.
As we anticipated above, the rank quality index $r$ is the average of
$r_{i,\alpha}$ over all the users in the data set:
 \begin{equation}
  r = \langle r_{i,\alpha} \rangle =
    \frac{\sum^{N}_{i=1} \sum^{M}_{\alpha=1} {r_{i,\alpha}} } { \sum^{N}_{i=1} \sum^{M}_{\alpha=1} {w_{i,\alpha}} }.
 \label{eq:rm}
 \end{equation}
Another validation measure testing the accuracy of the predictions is
the \textit{hitting rate}, $hit(L)$, i.e., for all the users, the ratio
between the number of collected objects included in the recommendation
list of length $L$, and the number of objects effectively collected
\emph{up to} the possible maximum value $N \cdot L$.  According to
these definitions, a good recommendation method should minimize the
value of $r$ and maximize the value of $hit(20)$.

For each data set, we considered $N_e = 20$ independent realizations
of the training set $T$, obtained by selecting uniformly
at random $90\%$ of the edges in the data set, we constructed the
recommendation list induced by each similarity measure, and computed
the value of the rank quality index $r$ and of the hitting rate
$hit(20)$. In the following we report the average values of $r$ and $hit(20)$ and their
associated statistical errors (the standard deviations of the means), respectively denoted by $\avg{r}$, $\avg{hit(20)}$, $\sigma_{\avg{r}}$ and
$\sigma_{\avg{hit(20)}}$. The mean values $\avg{r}$ and
$\avg{hit(20)}$ obtained over the $N_e=20$ different
realizations are shown in Table~\ref{table:r}.

\begin{table}[ht]

\caption{Average rank quality index $\avg{r}$ and hitting
  rate $\avg{hit(20)}$ for the different recommendation
  methods on each of the three data sets. The mean is computed on
  $N_e=20$ different realizations. The standard deviation of
  the mean values is given in parenthesis. For each database we
  highlight in boldface the best result.} 
\begin{tabular}{l c c c c} 
\hline\hline
\emph{MovieLens  } & $\avg{r}$ & $\sigma_{\avg{r}}$ & $\avg{hit(20)}$ & $\sigma_{\avg{hit(20)}}$ \\ [0.2ex] 
\hline
GRM  &  0.13821 & (0.00038)  & 0.1928 & (0.0041) \\  
\hline
 CF   &  0.11882 & (0.00037)  & 0.2364 & (0.0010) \\
 NBI  & {\bf 0.10514} & (0.00028)  & 0.2732 & (0.0010) \\
 MDW &  0.10563 & (0.00022)  &{\bf 0.2766}& (0.0010) \\
    BP &  0.10728 & (0.00032)  & 0.2708 & (0.0009) \\
    J &  0.11442 & (0.00035)  & 0.2568 & (0.0010) \\

\hline\hline
\emph{JesterJokes} &      &    \\ [0.2ex]
\hline
 GRM  &  0.30332 & (0.00049)  & 0.6160 & (0.0006) \\  
\hline
 CF   &  0.28718 & (0.00051)  & 0.6712 & (0.0008) \\
 NBI  &  0.28422 & (0.00045)  & 0.6775 & (0.0012) \\
 MDW &  0.28087 & (0.00037)  & {\bf 0.6806} & (0.0010) \\
    BP &{\bf 0.23795} & (0.00052)  & 0.6653 & (0.0012) \\
    J &  0.28549 & (0.00049)  & 0.6716 & (0.0014) \\

\hline\hline \emph{Fine Foods}  &    &   \\ [0.2ex] \hline
 GRM  &  0.22263 & (0.00073)  & 0.0891 & (0.0004) \\  
\hline
 CF   &  0.01458 & (0.00021)  & 0.7000 & (0.0014) \\
 NBI  &  0.01230 & (0.00012)  &{\bf 0.7402} & (0.0013) \\
 MDW &  0.01304 & (0.00017)  & 0.7293 & (0.0005) \\
    BP &  {\bf 0.01173} & (0.00010)  & 0.5777 & (0.0009) \\
    J &  0.01534 & (0.00013)  & 0.6990 & (0.0009) \\
\hline
\end{tabular}
\label{table:r}
\end{table}

By analyzing the results summarized in Table~\ref{table:r} we see that
the best results are obtained by different methods in different
databases. Moreover the two indicators $\langle r \rangle$ and
$\langle hit(20) \rangle$ always single out a different method as the
best one. However, an overall analysis shows that the best
recommendation methods are NBI (the best method according to $\langle
r \rangle$ in the MovieLens database and the best method according to
$\langle hit(20) \rangle$ in the Fine Foods database), MDW (the best
method according to to $\langle hit(20) \rangle$ in the MovieLens and
Fine Foods databases), and BP (the best method according to $\langle r
\rangle$ in the Jester Jokes and in the Fine Foods databases). They
clearly overcome the results obtained by GRM, CF, and Jaccard (J).

\section {Hybrid object-user methods} \label{Sec:Specular}

The most important difference between the recommendation methods
compared in Table~\ref{table:r} is that while NBI is based on a
definition of similarity among objects, all the other methods make use
of similarity measures defined between users.

In general, it is possible to define a transformation rule to obtain a
similarity score between users starting from a similarity between
objects, and viceversa. In fact, the similarity between objects
$s^{\rm o}_{i,j}$ can be obtained from the similarity between users by
appropriately swapping Latin indexes with Greek ones, and quantities
defined for users with the analogous ones defined for objects:
\begin{eqnarray}
s^{\rm u}_{i,j} & \leftrightarrow & s^{\rm o}_{\alpha,\beta} \nonumber \\
i,j & \leftrightarrow & \alpha,\beta \nonumber \\
N & \leftrightarrow & M.
 \label{spec}
\end{eqnarray}
The transformation rule is valid in both directions from user to objects and from objects to users.

We propose to define new recommendation scores by using the dual
similarity measures obtained with the above defined
transformation. For example, the recommendation value, which is the
dual of Eq.~(\ref{eq:vu}) and is valid for objects instead of users,
is obtained as:
 \begin{equation}
  v^{\rm o}_{i,\alpha} = \frac{\sum^{M}_{\beta=1, \beta\neq \alpha} {s^{\rm o}_{\alpha,\beta} \cdot a_{i,\beta}}}
 {\sum^{M}_{\beta=1, \beta\neq \alpha}{\left | s^{\rm o}_{\alpha,\beta}\right |} },
 \label{eq:vo}
 \end{equation}
whereas the dual recommendation score of the NBI algorithm is
 \begin{equation}
  f^{\rm u}_{i,\alpha} = \sum^{N}_{l=1} {s^{\rm u, NBI}_{i,l} \cdot a_{l,\alpha}}.
 \label{eq:fu}
 \end{equation}
It is interesting to note that according to the definition of the NBI we have
 \begin{equation}
  f^{\rm o}_{i,\alpha} = f^{\rm u}_{i,\alpha} = f_{i,\alpha}.
 \label{eq:f}
 \end{equation}
This relation can be verified by replacing $s^{\rm o,
  NBI}_{\alpha,\beta}$ (Eq.~\ref{s-NBI}) with $s^{\rm u, NBI}_{i,l} =
\frac{1}{k_l} \sum^{M}_{\beta=1} \frac{a_{i,\beta} a_{l,\beta}}
     {k_{\beta}}$ into the equations (\ref{eq:fo}) and (\ref{eq:fu}),
     respectively.
Hence, NBI is invariant under the transformation rules of
Eq.~(\ref{spec}). It is interesting to investigate how the duality
user/object similarity affects the quality of recommendation. To this
aim, we propose to define a recommendation value
$v_{i,\alpha}({\lambda})$ which is the result the convex combination of
the two recommendation values $v_{i,\alpha}^{\rm u}$ and
$v_{i,\alpha}^{\rm o}$ obtained from the similarity between users and
between objects, respectively. In formula:
 \begin{equation}
 v_{i,\alpha}(\lambda) = (1-\lambda) v^{\rm u}_{i,\alpha} + \lambda v^{\rm o}_{i,\alpha},
 \label{v-sym}
 \end{equation}
where the relative weight of the user and object recommendation values
is controlled by the parameter $\lambda\in [0,1]$, so that when
$\lambda=0$ we recover the recommendation score induced by the
similarity between users, while for $\lambda=1$ we have the
recommendantion score corresponding to the similarity between objects.
Our hypothesis, which is validated in the following, is that better
recommendations can be obtained by appropriately tuning the value of $\lambda$.

\begin{figure}
\includegraphics[angle=-90,width=8cm]{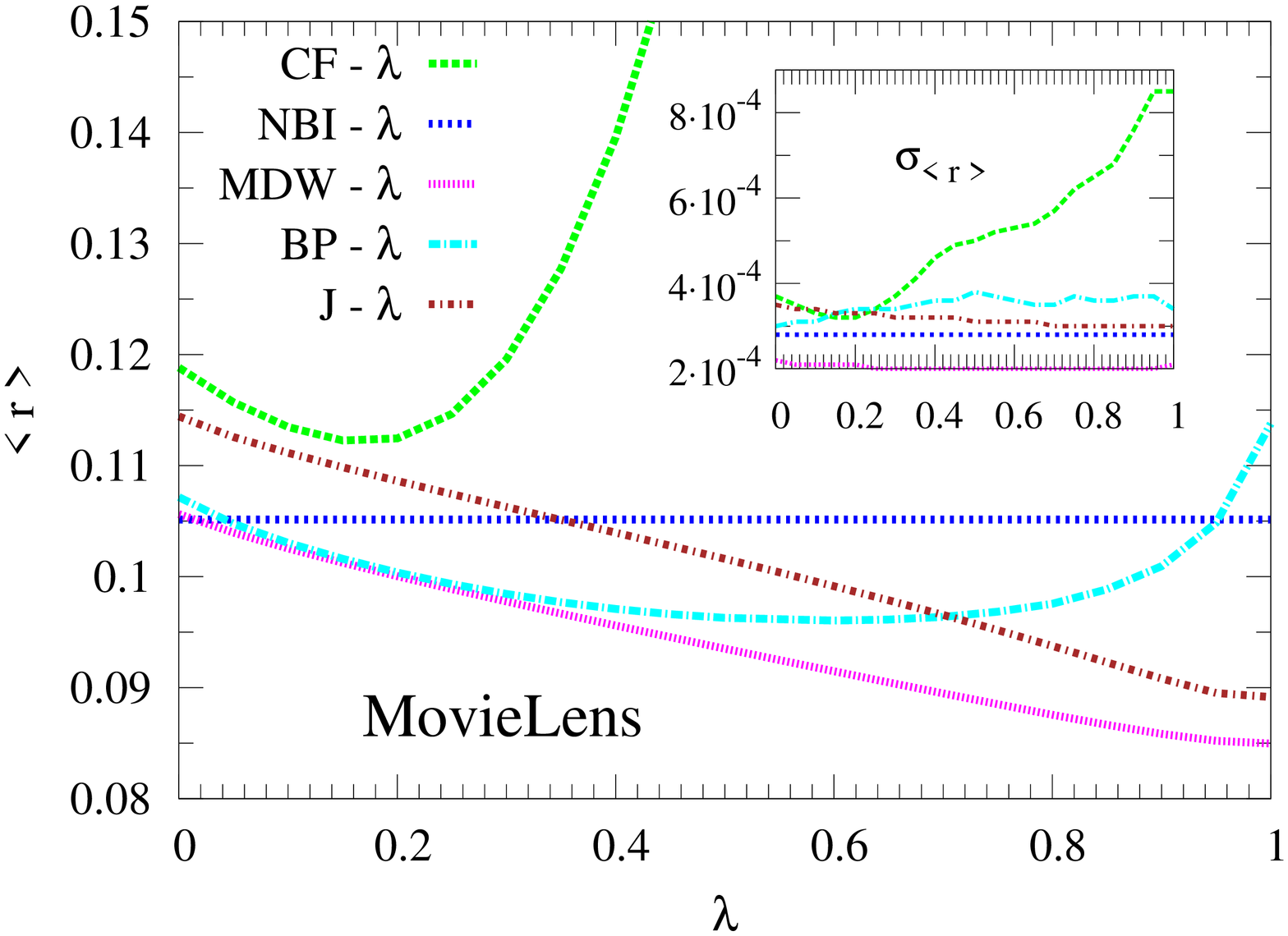}
\includegraphics[angle=-90,width=8cm]{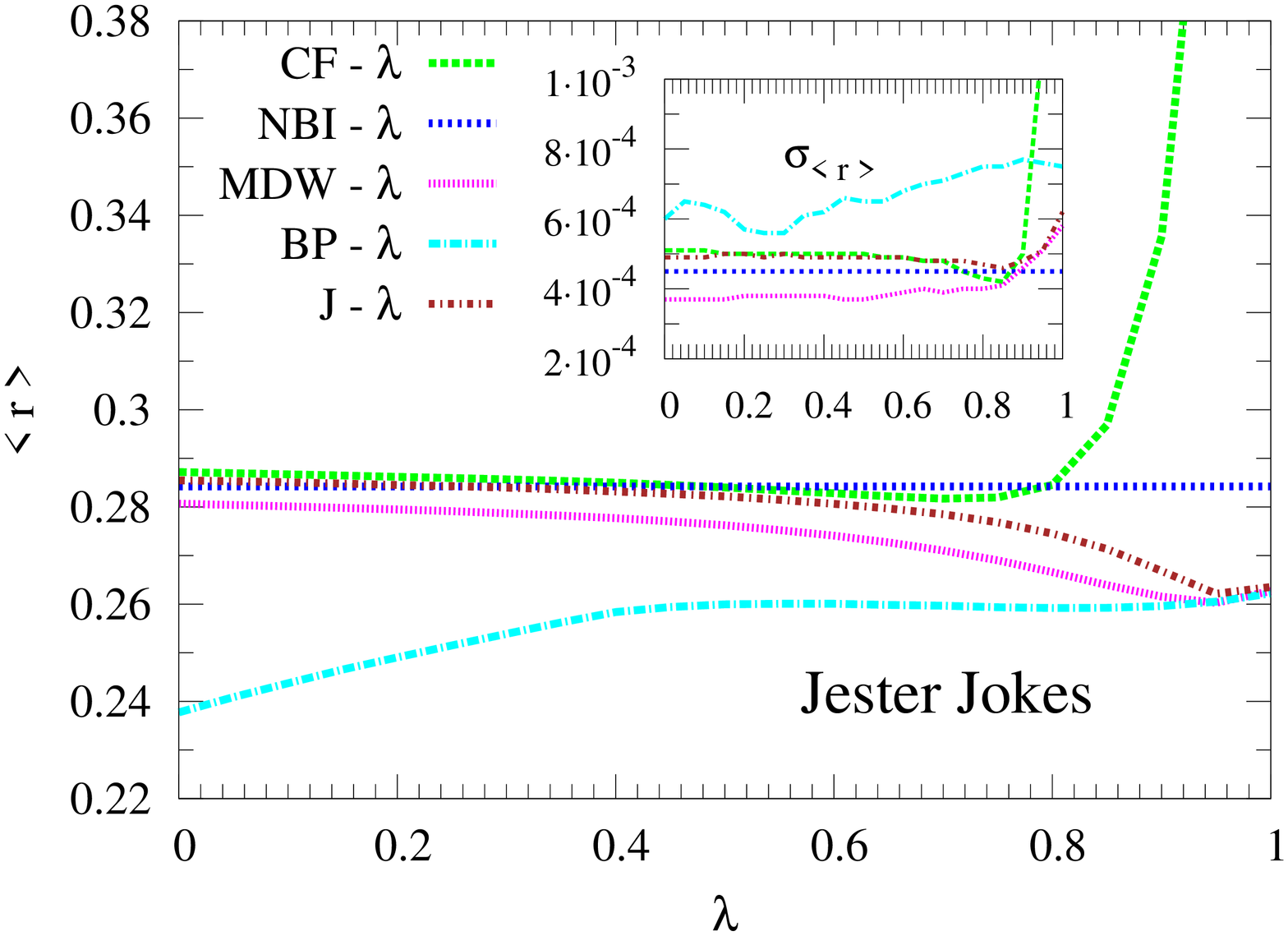}
\includegraphics[angle=-90,width=8cm]{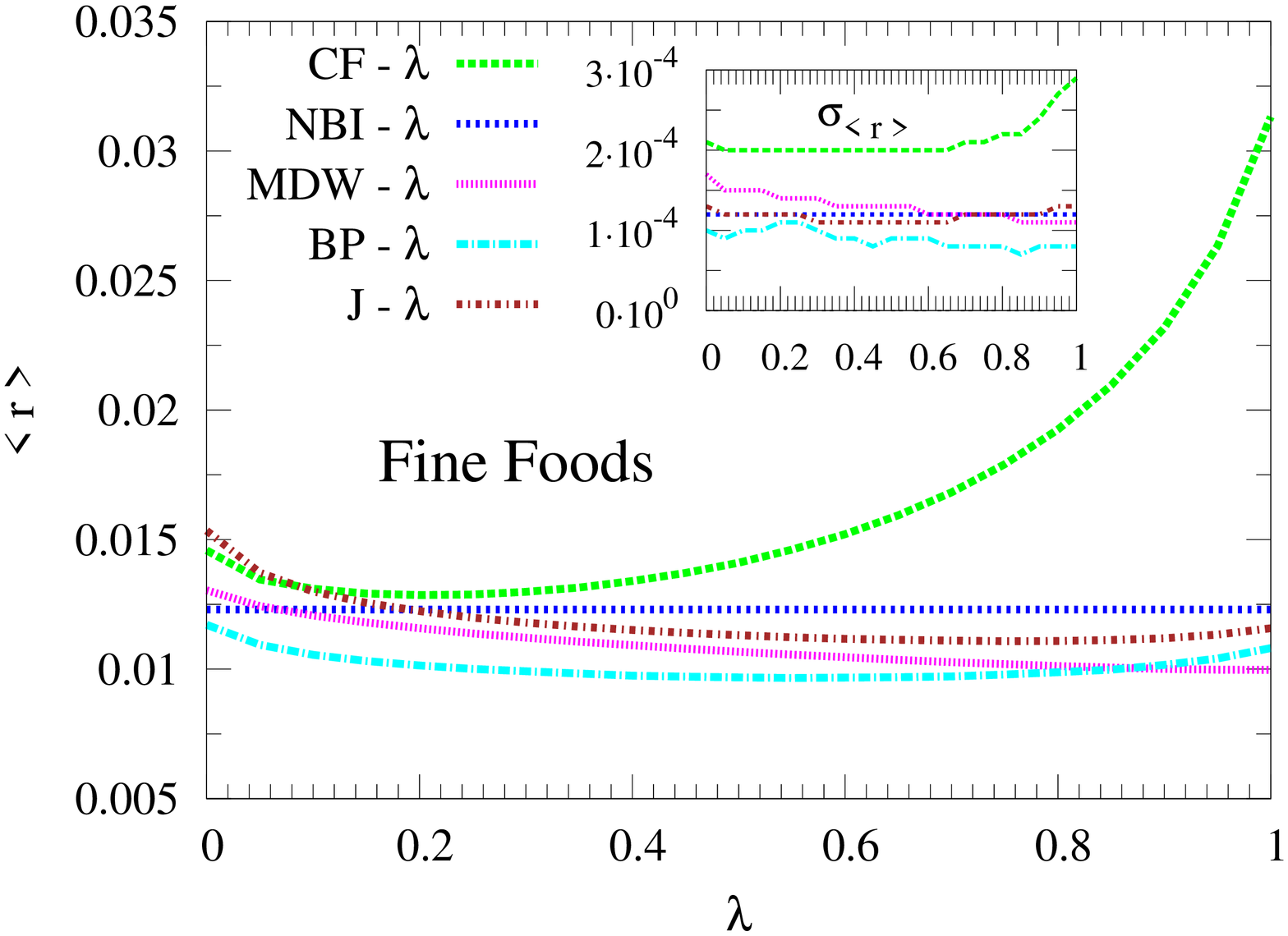}
\caption{(Color online) Average rank quality index $\langle r \rangle$ for the
different recommendation methods as a function of the user/object parameter $\lambda$.
 The insets report the corresponding standard errors of the mean.}
 \label{fig:sym}
\end{figure}

The mean values $\langle r \rangle$ for different recommendation
methods are reported in Fig.~\ref{fig:sym}, where the three panels
show the results obtained in the three data sets. It is worth noting
that the NBI algorithm is independent of $\lambda$.  In fact, by using
Eq.~(\ref{eq:f}) one verifies that
 \begin{equation}
 f_{i,\alpha} (\lambda) = (1-\lambda) f^{\rm u}_{i,\alpha} + \lambda
f^{\rm o}_{i,\alpha} = f_{i,\alpha}.
 \label{f-sym}
 \end{equation}

In Fig.~\ref{fig:sym} we notice that the CF recommendation method
performs poorly for almost all the values of $\lambda$, in all the
three data sets. In the case of MovieLens, three recommendation
methods (MDW, NBI and BP) perform in a similar way when only the user
similarity measure is taken into account ($\lambda=0$) as we already
noticed in the results summarized in Table~\ref{table:r}. On the other
hand, for $\lambda = 1$, i.e., when only the object similarity measure
is taken into account, the MDW method performs better than the
others. In the case of the Fine Foods data set, the BP similarity
performs slightly better than the others for $\lambda=0$ and for a
relatively large range of $\lambda$ values. When $\lambda = 1$, the
recommendation with the MDW measure performs slightly better. Finally, in the Jester Jokes data set the BP similarity clearly outperforms all the others when $\lambda = 0$, while for
$\lambda=1$ all the methods provide similar results, with the only
exception of the CF recommendation, whose performance is much worse.

The richness of profiles observed in Fig.~\ref{fig:sym} suggests that
the performance of a recommendation method depends  both on the
specific database and on the specific linear combination of user and
object recommendation values adopted. Quite often, the best
recommendation is not the one corresponding to $\lambda = 0$ or
$\lambda = 1$. Some methods perform better at the user limit
($\lambda=0$), others at the object limit ($\lambda=1$), some of them
for an intermediate value of $\lambda$. Moreover, the specific shape of $\avg{r}$
as a function of $\lambda$ actually depends on the database. We would like to stress two interesting
aspects of these results. First, some methods exhibit a convex profile
of $\langle r \rangle$ as a function of $\lambda$, where the minimum
indicates the best linear combination of user and object
recommendation values.  Second, the variability of the values of
$\langle r \rangle$ obtained by different recommendation systems is
much higher for $\lambda = 1$ than for $\lambda=0$.

\section {Impact of randomness}
\label{Randomization}

In this Section we analyse the robustness of recommendation systems
against the presence of different sources of noise in the data sets.
We consider three different kinds of randomization. In the first scenario we
add a certain amount of random edges to the bipartite graph, mimicking
erroneously reported user selections. In the second case we rewire a
given percentage of the edges of the bipartite network by maintaining
the degree of users unaltered (while the degree distribution of object
is in general modified). Finally, in the third case we rewire a
fraction of the edges of the graph by maintaining unaltered both the
user and object degree distribution.

\begin{figure}
\includegraphics[angle=-90,width=8.0cm]{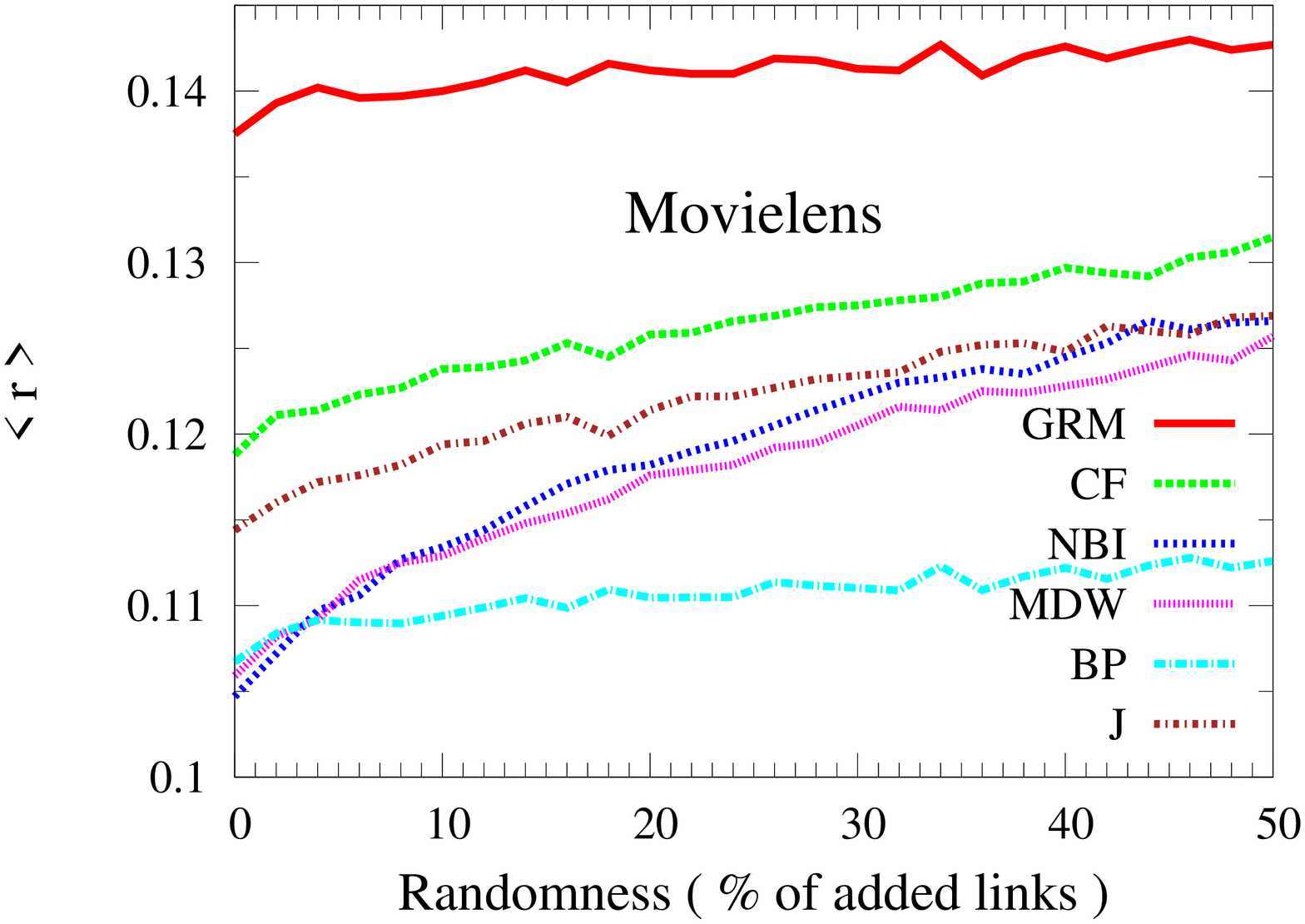}
\includegraphics[angle=-90,width=8.0cm]{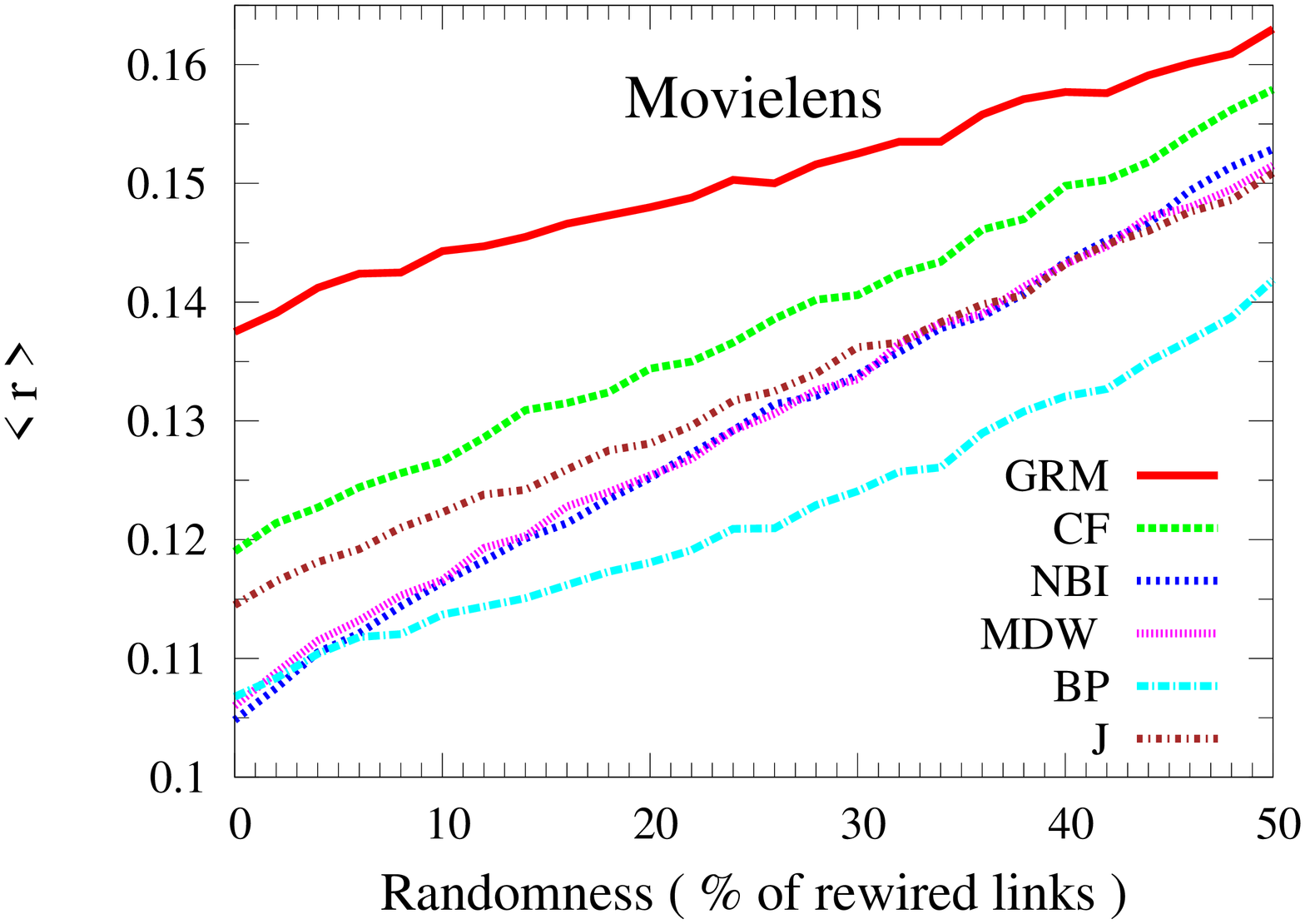}
\includegraphics[angle=-90,width=8.0cm]{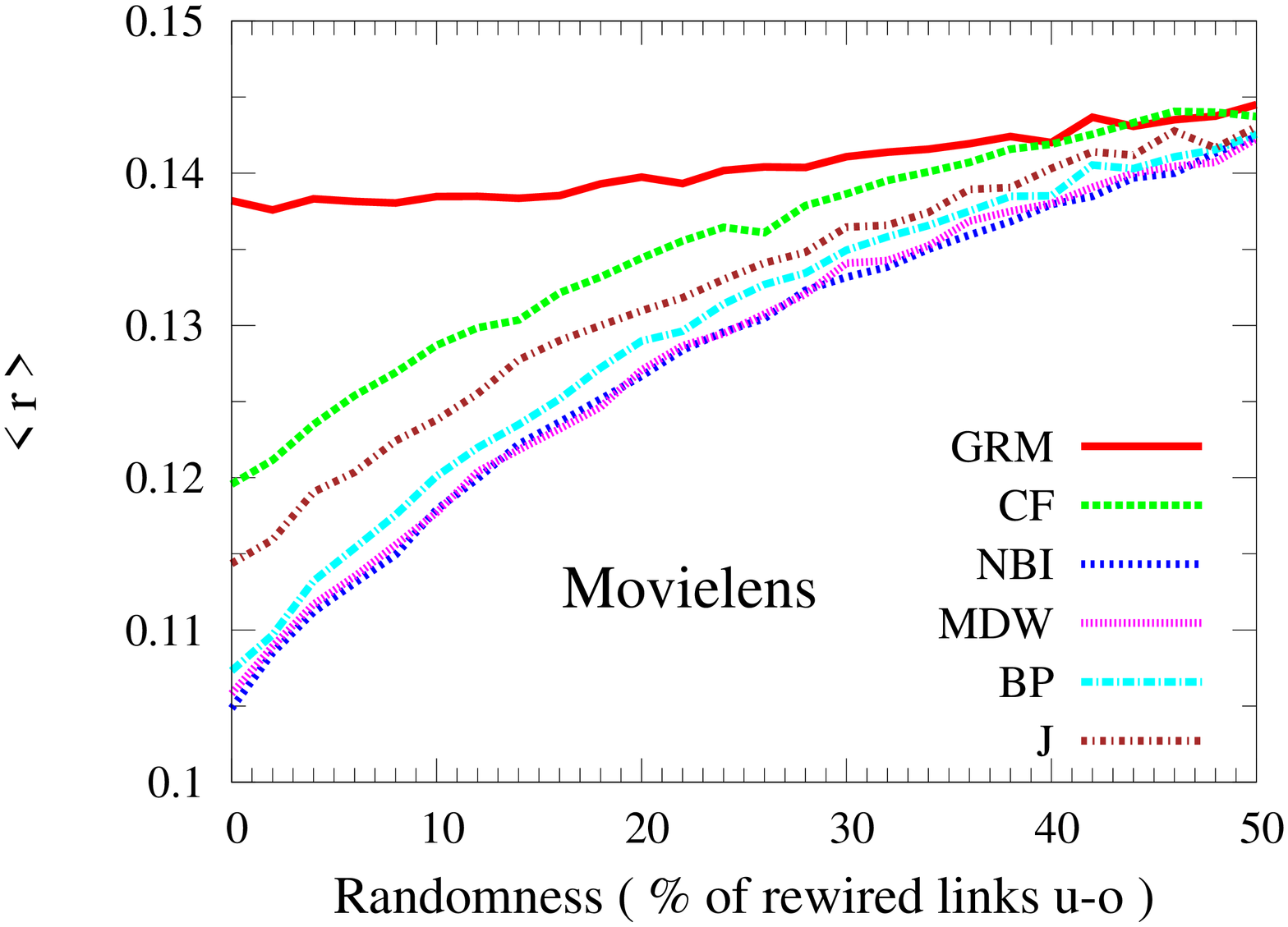}
\caption{(Color online) Mean validation values as a function of noise
  in the MovieLens database. The three panels correspond,
  respectively, to the addition of edges at random (top) and to edge
  rewirings which maintain unaltered only the user degree sequence
  (middle) or both the user and object degree sequences (bottom).}
 \label{fig:m_rand}
\end{figure}

For the sake of simplicity, we show the results obtained for the three
randomizing methods only for the MovieLens database.  In
Fig.~\ref{fig:m_rand} we show the average rank quality index for the
different methods with $\lambda=0$ as a function of the percentage of edges randomly added
or rewired. As expected, $\avg{r}$ is an increasing function of the
percentage of noise, signalling a degradation of the recommendation
performance. However, the actual profile of $\avg{r}$ depends on the
specific recommendation method used. In fact, several curves crosses
at different values of the induced randomness. This is clearly
observed for the first and second kinds of randomization.

\begin{figure}
\includegraphics[angle=-90,width=8.0cm]{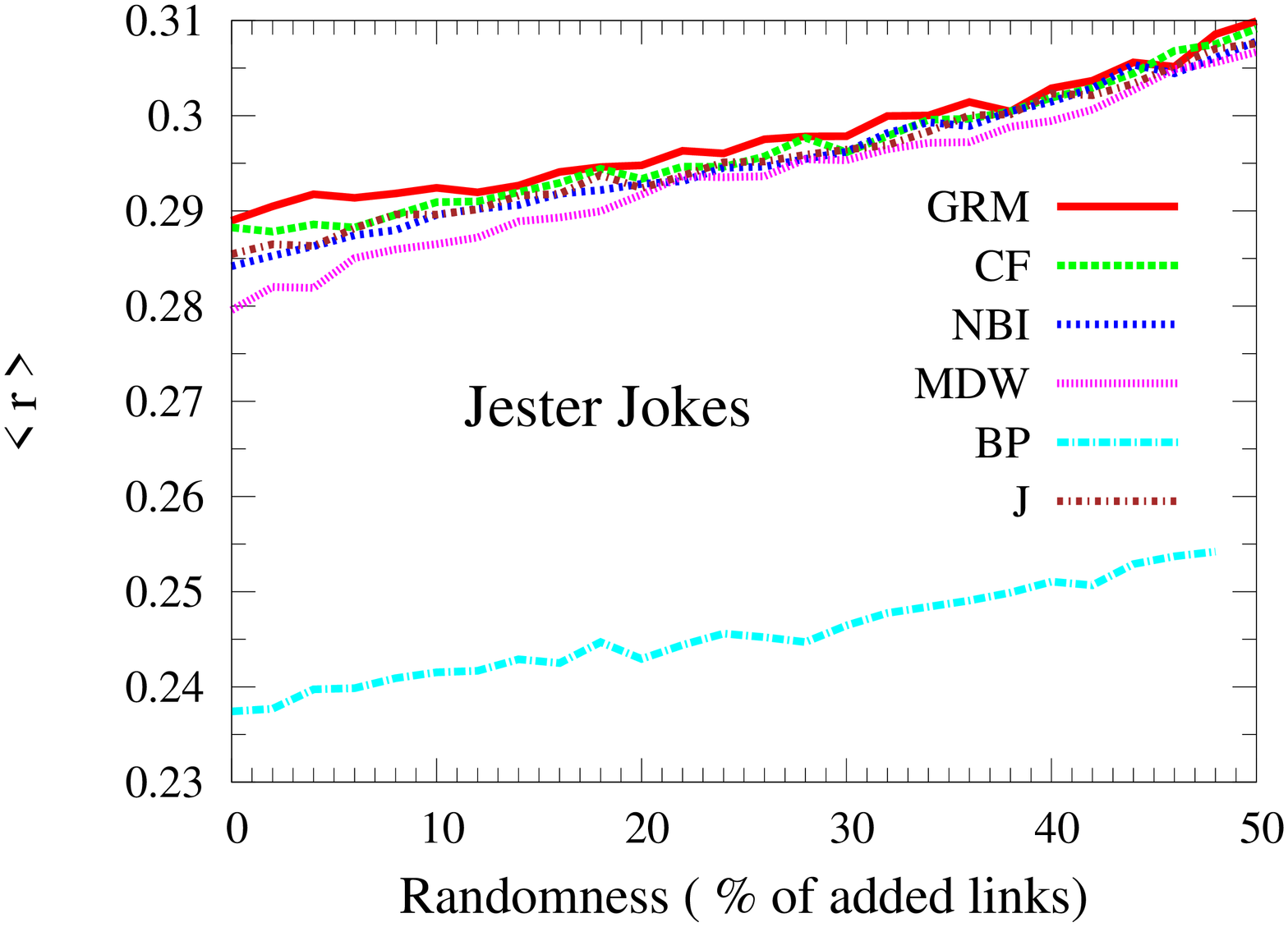}
\includegraphics[angle=-90,width=8.0cm]{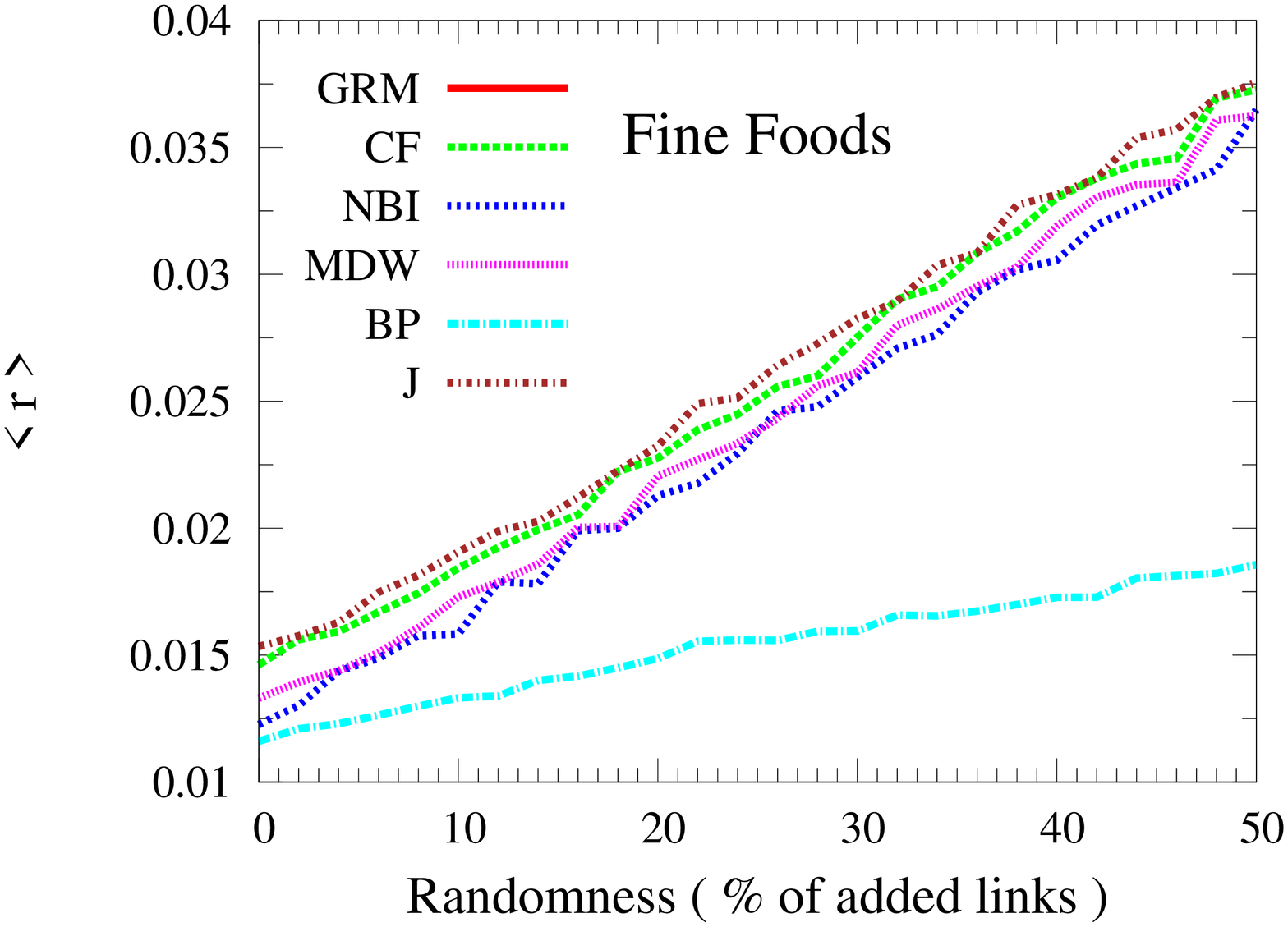}
\caption{(Color online) Mean validation values for some different
  methods as a function of the randomness with the Jester jokes (top
  panel) and Fine Foods (bottom panel) databases. }
 \label{fig:j_rand}
\end{figure}

We performed the same analysis also on the Jester Jokes and Fine Foods
databases, and we report in Fig.~\ref{fig:j_rand} the results
corresponding to the first type of randomization (addition of a fraction of
random edges). The results show a prominent role of the BP similarity
measure, which seems the most robust in dealing with noisy data sets.

Our findings suggest that the BP similarity measure is a good
candidate to provide good and robust recommendations in databases
where there is a high degree of uncertainty about the validity of
records. In fact, while the use of the BP similarity does not give
substantially better recommendation prediction in databases like
MovieLens and Fine Foods, its performance is consistently higher in
the case of Jester Jokes.


\section {Conclusions}

We have considered three real-world users/items bipartite networks, we
have investigated the performance of several traditional
recommendation methods recently presented in the literature, and we
have proposed two new similarity measures which take into account the
heterogeneity of users and objects degrees. We showed that these two
new similarity indexes can outperform traditional recommendation
systems in most of the cases, even if there is a clear dependence of
the results on the structural characteristics of the data set under
study.

Then, we focused on hybrid recommendation systems based on the convex
combination of the recommendation scores induced by the similarity
between users and objects, parametrised by a coefficient $\lambda$. We
showed that different outcomes can be obtained in personalized
recommendation methods by using similarity between users, or between objects,
or a combination of the two. In some cases,  the quality of recommendation 
as measured by the average rank quality index $r$ is a convex function of 
the parameter $\lambda$. This means that the combination of different 
recommendation scores might actually provide better performance with 
respect than the employment of user or object similarities alone and, 
more importantly, that depending on the data set at hand, the quality of recommendation can be actually
optimised through an appropriate tuning of $\lambda$. Conversely, 
for some similarity measures we observed a monotonically decreasing dependence of 
$\avg{r}$ on $\lambda$, so that the best recommendation is obtained by using 
an object-based similarity.
We finally investigated the robustness of recommendation systems to
the addition and rewiring of edges, and the results suggested that the Binary Person correlation similarity can consistently outperform other similarity measures in
noisy data sets.

Although we do not observe a specific recommendation method
outperforming all the others in all conditions and for all the data
sets considered, it seems that recommendations based on MDW and BP
are able to produce better results than those using other similarity
measures.  However, our results show that the performance of the
recommendation methods depends on both the specific investigated
database and on the way similarities between users and objects  are
used to derive recommendation scores.

\begin{acknowledgments}
This work is partially supported by the EPSRC project GALE
EP/K020633/1. This research utilised Queen Mary's MidPlus
computational facilities, supported by QMUL Research-IT and funded
by EPSRC grant EP/K000128/1.

\end{acknowledgments}


\begin{thebibliography}{99}

\bibitem{2005IEEEAdomavicius} G. Adomavicius and A. Tuzhilin, \textit{IEEE} {\bf 17} 734 (2005).

\bibitem{2012PRLu} L. L\"{u}, M. Medo, C.-H. Yeung, Y.-C. Zhang, and Z.-K. Zhang, \textit{Phys. Rep.} {\bf 519} 1 (2012).

\bibitem{2001Sarwar} B. Sarwar, G. Karypis, J. Konstan, and J. Riedl. Item-based collaborative filtering. Proc. Int. Conf. WWW10, ACM 1-58113-348-0/01/0005, 285 (2001) 

\bibitem{1992ACMGoldberg} D. Goldberg, D. Nichols, B.M. Oki, D. Terry, Commun. ACM {\bf 35} 61 (1992) 

\bibitem{2007Schafer} J.B. Schafer, D. Frankowski, J. Herlocker, S. Sen, \textit{Collaborative filtering recommender systems. In: The adaptive web} Springer 291 (2007). 

\bibitem{2007PRLZhang} Y.-C. Zhang, M. Blatter, and Y.-K. Yu, \textit{Phys. Rev. Lett.} {\bf 99} 154301 (2007).

\bibitem{2009NJPZhou} T. Zhou, R.Q. Su, R.R. Liu, L.L. Jiang, B.H. Wang, and Y.-C. Zhang, \textit{New J. of Phys.} {\bf 11} 123008 (2009).

\bibitem{2007PREZhou} T. Zhou, J. Rien, M. Medo, and Y.-C. Zhang, \textit{Phys. Rev. E} {\bf 76} 046115 (2007).

\bibitem{2011PRELu} L.  L\"{u}, W. Liu, \textit{Phys. Rev. E} {\bf 83} 066119 (2011)

\bibitem{2009PNASZhou} T. Zhou, Z. Kuscsik, J.-G. Liu, M. Medo, J.R. Wakeling, and Y.-C. Zhang, \textit{PNAS} {\bf 107} 4511 (2010).

\bibitem{2013PONEQiu} T.Q. Qiu, Z.-K Zhang, and G. Chen,\textit{PloSONE} {\bf 8} 1 (2013).

\bibitem{2014arXivZhu} X. Zhu, H. Tian, and S. Cai, arXiv:1405.4095v1 [cs.IR] (2014).

\bibitem{1895PRSLpearson} K. Pearson, \textit{Proceedings of the Royal Society of London} {\bf 58} 240 (1895).

\bibitem{2011PALu} L. L\"{u}, T. Zhou, \textit{Physica A} {\bf 390} 1150 (2011).

\bibitem{1912NPjaccard} P. Jaccard, \textit{Bull. Soc. Vaud. Sci. Nat.} {\bf 37} 547 (1901).

\bibitem{2009PALiu} R.-R. Liu, C.-X. Jia, T. Zhou, D. Sun, B.-H. Wang, \textit{Physica A} {\bf 388} 462 (2009).

\bibitem{2012EPJBGuo} Q. Guo, R. Leng, K. Shi, and J.G. Liu, \textit{Eur. Phys. J. B} {\bf 85}: 286 (2012)

\bibitem{2011PoneAGing} M. Tumminello, S. Miccich\`e, L.J. Dominguez, G. Lamura, M.G.
Melchiorre, M. Barbagallo, and R.N. Mantegna, \textit{PlosONE} {\bf 6}(9) e23377 (2011)

\bibitem{2014Hatzopoulos} V. Hatzopoulos, G. Iori, R.N. Mantegna, S. Miccich\`e, and M. Tumminello: \textit{Quantitative Finance} DOI:10.1080/14697688.2014.969889 (2014)

\end{thebibliography}
\end{document}